\documentclass[10pt,conference]{IEEEtran}

\usepackage{cite}
\usepackage{amsmath,amssymb,amsfonts}
\usepackage{algorithmic}
\usepackage{graphicx}
\usepackage{textcomp}
\usepackage{xcolor}
\usepackage{url}
\usepackage{microtype}
\usepackage{caption}

\begin{document}

\title{\textbf{\Large Tackling Software Architecture Erosion: \\ 
Joint Architecture and Implementation Repairing by a Knowledge-based Approach\\[-1.5ex]}}

\IEEEoverridecommandlockouts
\IEEEpubid{\begin{minipage}{\textwidth}\ \\[11pt] \centering
\copyright 2021 IEEE. Personal use of this material is permitted. Permission from IEEE must be obtained for all other uses, in any current or future media, including reprinting/republishing this material for advertising or promotional purposes, creating new collective works, for resale or redistribution to servers or lists, or reuse of any copyrighted component of this work in other works.
\end{minipage}}
\IEEEpubidadjcol

\author{
\IEEEauthorblockN{Christoph Knieke, Andreas Rausch, Mirco Schindler}
\IEEEauthorblockA{
\textit{Clausthal University of Technology}\\
Clausthal-Zellerfeld, Germany \\
Email: firstname.surname@tu-clausthal.de}\\
}
\maketitle


\begin{abstract}
Architecture erosion is a big challenge in modern architectures leading to a deterioration of the quality properties of these systems. 
Today, no comprehensive approach for regaining architecture consistency in eroded software systems exists and architecture consistency is essentially achieved by repairing the implementation level only. 
In this paper, we propose a novel approach enabling a joint architecture and implementation repairing for tackling software architecture erosion. 
By using a holistic view on violation causes and suitable repair actions in combination with learning mechanisms we build up a system specific knowledge-base improving accuracy and efficiency in consolidation of architecture and implementation over time. 

\end{abstract}

\maketitle

\section{Introduction}

The term (software) architecture erosion refers to the process of continuous divergence between the intended software architecture of a system and its refinement or implementation \cite{VanGurp2002}. The causes of architecture erosion are manifold, such as changes to the code that are not consistent with the architecture in order to implement bug fixes or additional requirements \cite{Lindvall2008Bridging}. The deviation of the implementation of the system from the intended architecture generally leads to a degradation of the quality properties of the system such as maintainability, extensibility, or reuse in the medium term \cite{Knodel2007Comparison}. In the long term, untreated architecture erosion leads to hard-to-maintain systems that must be replaced by new developments \cite{Sarkar2009Modularization}.
Despite the ubiquity of this problem of (software) architecture erosion, no comprehensive approach to regaining architecture consistency in degraded software systems exists today \cite{ali2018architecture, de2012controlling}. 

The software architecture $A$ of a system can be understood as a set $\Phi$ of predicate logic statements about the structure and behavior of the system. 
For a given tuple of the architecture $A$ and the corresponding implementation $S$, we have consistency between $A$ and $S$, if and only if the implementation $S$ (where by the set $S$, we mean the set of all artifacts of that implementation) satisfies the predicate logic statements of $\Phi$ (denoted as $S \models \Phi$) and thus conforms to the architecture specification. In an eroded system, the implementation is not architecture conformant (\mbox{$S\not\models \Phi$}), i.e., there are architecture violations.

If consistency between architecture $A$ and implementation $S$ is now to be regained, either $S$ must be converted into an architecture-compliant implementation $S'$ s.t. $S' \models \Phi$ (Implementation Repairing) or $A$ must be adapted accordingly, i.e., we have to change $\Phi$ to $\Phi'$ s.t. $S \models \Phi'$ (Architecture Repairing). In a previous work, 
we have already proposed an approach for Implementation Repairing \cite{mair2014}. Now, we introduce a novel approach where we consider inconsistent tuples of $A$ and $S$ and enabling the joint adaptation of $S$ and $A$.

For the adaptation, repair actions are available, e.g., in the form of a set of refactorings. 
Determining an optimal sequence of repair actions requires an exhaustive search, which, however, must fail for realistic scenarios due to the size of the search space in combination with the computational complexity of architecture conformance \cite{mair2014}. Practicable solutions therefore limit the problem to at least one of the following: reducing the complexity of considered architectural concepts and violations, reducing the considered repair actions, and using heuristic search methods \cite{mair2014}.

\section{Proposed Methodology}

\begin{figure}[!h]
\centering
\includegraphics[width=1.0\columnwidth]{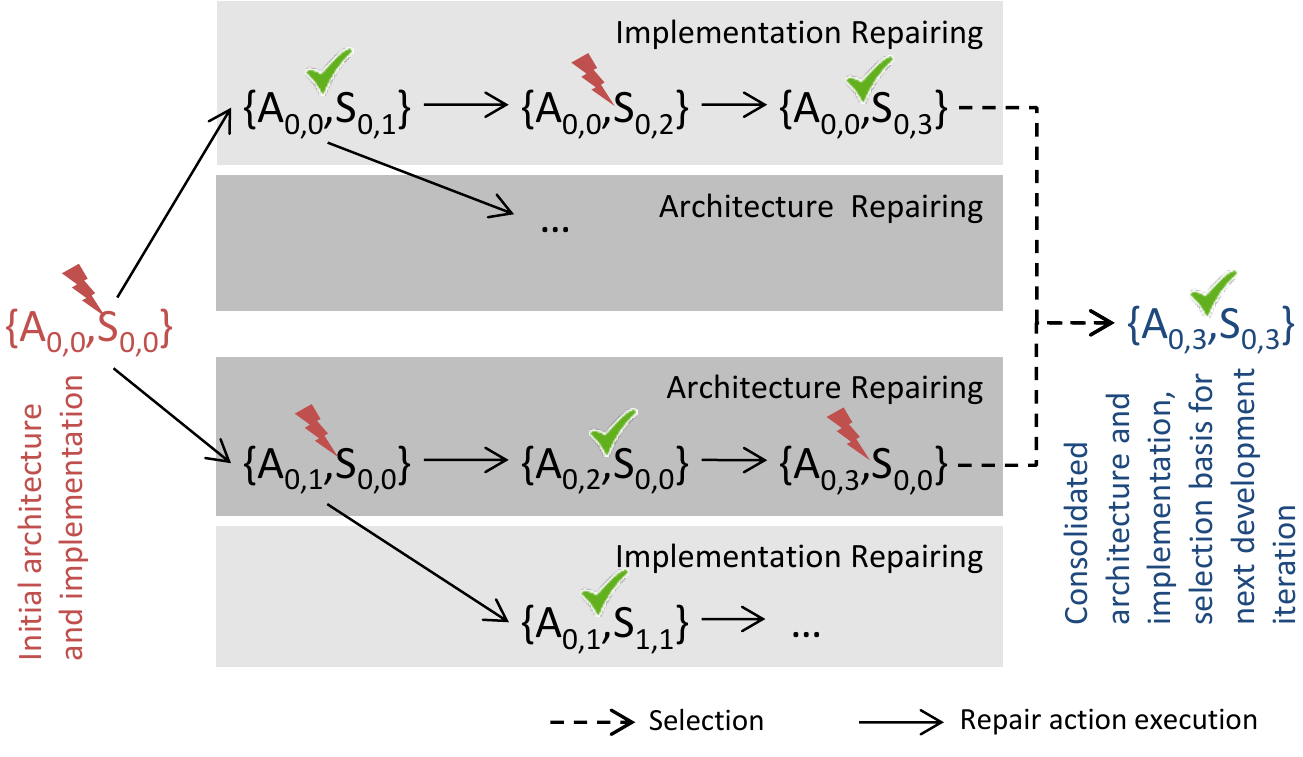}
\caption{Transfer to a consolidated architecture and implementation}
\label{fig:tuple}
\end{figure}

Figure~\ref{fig:tuple} shows possible paths in the transition of an inconsistent tuple of $A$ and $S$ to a consolidated architecture and implementation. Each edge corresponds at least to the execution of a repair action either on the implementation level or on the architecture level. This single modification need not lead to architecture and implementation being consistent with each other again, i.e., on the way to a consolidated architecture and implementation, repair actions may first lead to degradation. Furthermore, it may be more optimal to make changes to both the architecture and the implementation.

\begin{figure*}[!t]
\centering
\includegraphics[width=2.0\columnwidth]{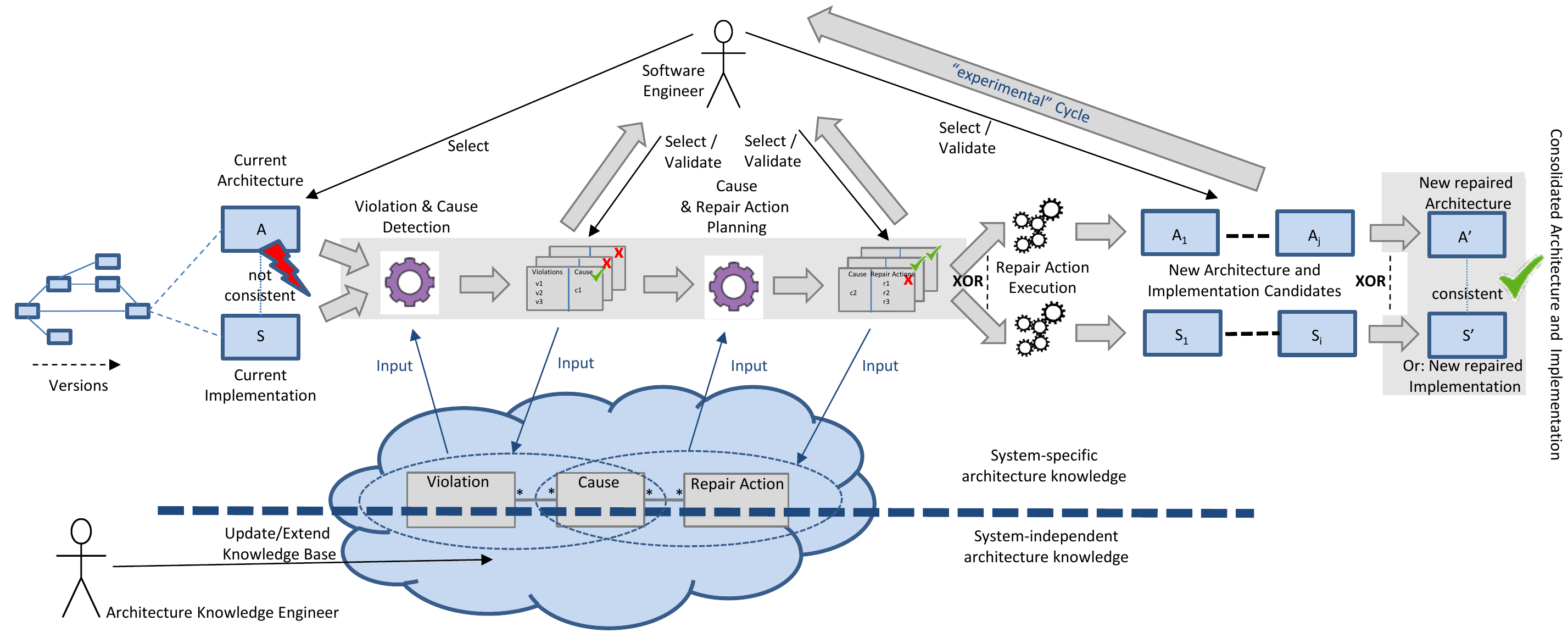}
\caption{Conceptual overview of the proposed approach}
\label{fig:approach}
\end{figure*}

This concept shown above is the foundation of our methodology (see Figure~\ref{fig:approach}). 
The solution approach is based on a managed creativity-experimentation cycle, where the software engineer is even encouraged to just try things out, in the sense of trial and error, but by recording this process, knowledge can be extracted and integrated into a knowledge base ("cloud" in Figure~\ref{fig:approach}). 

\newpage Thereby, future recommendations can be improved and it is possible to transfer the recommendations to other systems, s.t. in similar systems in the company proven solutions can be proposed. 
The generic causes, failure patterns, and repair actions from the system-independent architecture knowledge constitute the baseline. Learning is applied to refine this knowledge. In order to take into account the system-specific decisions and characteristics, system-specific knowledge is built up in particular. 

The use of refactorings or the introduction of established patterns, e.g., can lead to the fact that well-established object-oriented metrics have ”worse” results \cite{Burger2013}. Therefore, it is also essential to generate system-specific knowledge and tune the knowledge base for a concrete system.

The methodology starts with a not consistent tuple of $A$ and $S$ as input. The first step with the \textit{Violation \& Cause Detection} activity aims to inspect $A$ and $S$ and detecting inconsistencies between them. Combinations of multiple architecture violations and properties of the associated implementation fragments generally indicate a deeper problem behind these inconsistencies. To do this, information about individual instances of architectural violations is aggregated to create an overall picture (failure pattern) of the underlying causes of the discrepancies. The output of the \textit{Violation \& Cause Detection} activity is a set of pairs consisting of architectural violations and an underlying cause for those violations. The activity also takes as input the system-specific architecture knowledge with pairs of violations and their cause that have already been detected in the past. From this set of generated pairs, the software engineer selects one that he/she considers plausible and thus determines the input for the next activity.

Based on the identified and validated causes and the violations, it must be decided whether and how the current degradation can be repaired. The \textit{Cause \& Repair Action Planning} activity aims at recommending either the repair of $S$ with respect to $\Phi$ or the change of $\Phi$ with respect to  $S$ based on the identified causes of degradation. To do this, the activity determines a set of pairs consisting of cause and a sequence of repair actions for the selected pair of violations and cause. In addition, the input to this activity is the knowledge base and the sequences of successful repair actions for specific causes already learned there. 

In the \textit{Repair Action Execution} activity, repair actions are now applied in sequence. As a result, new intermediate versions of the repaired system or the repaired architecture are created. Now it is possible to start with the next repair action or to go back one or more steps if the result turns out to be unsatisfactory. 

It is possible that going through these activities once may not lead to an optimal result in every case. Thus, the approach explicitly provides, in the sense of backtracking, for re-evaluating decisions made and for the software engineer to retract them, e.g., to choose other repair actions or even other causes. The latter would also lead to the re-execution of the \textit{Cause \& Repair Action Planning} for the now favored cause. 


\section{Conclusion and Future Work}
We proposed a novel approach for tackling architecture erosion using learning mechanisms and a joint architecture and implementation repairing. 
As a future work, we plan to implement the sketched approach based on a set of ground truth architectures and to finally evaluate our approach.

\bibliographystyle{IEEEtran}
\bibliography{lit,MyCollection}

\end{document}